\documentclass[12pt]{article}

\usepackage{amsmath,amssymb,bbm}
\usepackage{epsfig}
\usepackage{slashed}
\usepackage{color}

\title{
\vspace{-2cm}
\vspace{3cm}
\bf \huge
Muon Magnetic Moment\\ and the Goldstone Boson Higgs\vspace{1cm}}
\date{}
\author{{Oleg Antipin\footnote{oleg.antipin@fi.infn.it}, Stefania De Curtis\footnote{stefania.decurtis@fi.infn.it}, Michele Redi\footnote{michele.redi@fi.infn.it}, Carlotta Sacco\footnote{carlottasacco@virgilio.it}}\\
[10mm]
\normalsize\itshape  INFN, Sezione di Firenze, Via G. Sansone, 1; I-50019 Sesto Fiorentino, Italy
}

\begin{document}
\maketitle
\begin{abstract}
\medskip
\noindent
We compute the correction to the muon  magnetic moment in theories where the Higgs is a
pseudo-Goldstone boson and leptons are partially composite. Using a general effective lagrangian we show that in some regions of parameters 
a sizable new physics contribution to the magnetic moment can be obtained from composite fermions that could explain the 3.5$\sigma$ experimental discrepancy from the Standard Model prediction. This effect depends on the derivative interactions of the Higgs that do not modify the coupling of the Higgs to leptons and it does not require
extremely light fermions, allowing to easily avoid LHC bounds. Our derivations can be in general applied to dipole operators in theories with Goldstone boson Higgs.
\end{abstract}

\newpage
\tableofcontents

\section{Introduction}

In this note we study new physics contributions to the muon anomalous magnetic moment in theories where
the Higgs is a Goldstone boson (GB) and leptons are partially composite, see \cite{review} for a review. 
These models are strongly motivated by the hierarchy problem because the Higgs boson, being a composite state, is not 
sensitive to scales much shorter than its size. This points to a scale of compositeness around TeV
that is being tested at the LHC.

Our main motivation here is the long standing muon magnetic moment anomaly
\begin{equation}
\Delta a_\mu=a_\mu^{exp}-a_\mu^{SM}= (2.8\pm 0.8) \times 10^{-9}
\label{anomaly}
\end{equation}
($a_\mu=(g_\mu-2)/2$) whose size suggests a new physics contribution of the order of the Standard Model (SM) electro-weak correction.
In renormalizable theories where SM fields mix with heavy leptons the contribution scales as
\begin{equation}
\Delta a_\mu \sim \frac {g_{\psi}^2}{(4\pi)^2} \frac {m_\mu^2}{\Lambda^2} 
\end{equation}
where $\Lambda$ is a new physics scale associated with the heavy fermions and $g_\psi$ their coupling to the Higgs.
At face value the effect is typically too small unless the fermions are as light as 200-300 GeV, 
in agreement with explicit models \cite{bib:strumia,bib:dermisek,bib:westhoff}.
In theories with GB Higgs new diagrams arise from the non-linearities of the theory demanded 
by the symmetries and also UV contributions from the composite sector dynamics are expected.
We wish to show that the size of these effects could account in certain regions of parameters for the anomaly (\ref{anomaly}),
compatibly with  bounds from flavor physics, LHC searches and electro-weak precision tests. 

\section{Partially Composite Muon}
\label{sec:2}

We work within the framework of composite Higgs models with partial compositeness.
The Higgs is a GB of some strongly coupled theory with global symmetry $G$ spontaneously broken 
to a subgroup $H$ at a scale $f>v$. For concreteness we will focus on the minimal models based on $SO(5)/SO(4)$
but our results can be extended to other patterns of symmetry breaking and different representations, see \cite{C2HDM}.
SM fermions are partially composite, mixing with states of equal quantum numbers under the SM gauge symmetries.

The lagrangian for the composite states can be described in the most general fashion using 
the CCWZ  formalism \cite{CCWZ}. We focus here on new composite fermions and do not include
vector resonances for simplicity and because they are typically heavier.  Composite states 
are classified according to their representation under the unbroken group. The most general lagrangian compatible with the symmetries can be constructed 
with the aid of the connections $e_\mu$ and $d_\mu$ by writing down all possible invariants under the unbroken group. 
The connections are explicitly reported in appendix \ref{appendixA} for the coset $SO(5)/SO(4)$. Elementary fields can be introduced 
assigning them to a representation of $SO(5)$ and writing the most general couplings to the composite 
states using the GB matrix $U$.

For concreteness we  study in detail the scenario where the  left and right chirality of the muon couple to composite 
fermions in the {\bf 5} of $SO(5)$ but we provide the tools for computing in general dipole  moments in models with GB Higgs. 
For the top quark, a model with the same structure can be found in \cite{bib:panico}, see also \cite{bib:nonminimal}. We refer to \cite{bib:panico} and appendix~\ref{appendixA} 
for details on the notation. We focus on a single generation and comment on the flavor structure in Sect.~\ref{discussion}.
The composite states decompose into a quadruplet and a singlet under $SO(4)$,
\begin{equation}\label{emb}
 {\bf 5}= {\bf 4}+{\bf 1}\ : \qquad \quad \psi_4=\frac{1}{\sqrt{2}}
\left( \begin{array}{c}
i(E_{-2}-N) \\
E_{-2}+N\\
i(E_{-1}+E)\\
E-E_{-1}
\end{array} \right)\,,~~~~~~~~~~~~~~\psi_1=\tilde{E}
\end{equation}
with lagrangian
\begin{eqnarray}
\label{comp5}
\mathcal{L}_{comp}&=&\overline{\psi}_4(i\,\displaystyle{\not} D-m_4)\, \psi_4+\overline{\psi}_1(i\, \displaystyle{\not} D\,-m_1)\, \psi_1 \nonumber \\
&+&i\, d_\mu^{\,\hat{a}}\, \left[\, c_L\,\overline{\psi}_{4L}^{\,\hat{a}}\,\gamma^\mu\, \psi_{1L}+ c_R\,\overline{\psi}_{4R}^{\,\hat{a}}\,\gamma^\mu\, \,\psi_{1R} +h.c.\right]
\end{eqnarray}
where
\begin{eqnarray}\label{dercov1}
D_\mu\psi_1 &=&[\partial_\mu +i\,g' B_\mu]\,\psi_1\nonumber \\
D_\mu\psi_4 &=&[\partial_\mu- i\,e_\mu\,+ i g' B_\mu]\, \psi_4\,\nonumber \\
d_\mu^{\hat{a}}&=&\frac{\sqrt{2}}f D_\mu \pi^{\hat{a}}+\dots
\,,\label{dercov4}
\end{eqnarray}
$g'$ and $B_\mu$ are the SM hypercharge coupling and field and $\pi^{\hat{a}}$ are the four components of the Higgs doublet. 
The second line in (\ref{comp5}) contains the leading derivative interactions of the Higgs with 
the fermions (controlled by the symmetry breaking scale $f$) that are characteristic of GB theories. 
These will play a crucial role in the computation of $\Delta a_\mu$. 
Based on general power counting arguments we assume $c_{L,R}$ to be of order one \cite{bib:wulzer}.
The derivative couplings  are in general complex unless the composite sector respects $CP$. Moreover $c_L=c_R$ if parity is preserved.  

The mixing with the elementary fermions is given by,
\begin{eqnarray}\label{mixings}
-\mathcal{L}_{mixing}&=& y_{L_4}\, f\,(\bar{l}_L^{\,\mathbf{5}})^IU_{I{\hat a}}\,\psi_4^{\hat a} +y_{L_1}\, f\,(\bar{l}_L^{\,\mathbf{5}})^IU_{I5}\,\psi_1 +\nonumber \\
&+&y_{R_4}^*\, f\,(\bar{\mu}_R^{\mathbf{5}})^IU_{I{\hat a}}\,\psi_4^{\hat a} + y_{R_1}^*\, f\,(\bar{\mu}_R^{\mathbf{5}})^IU_{I5}\,\psi_1 +h.c.
\end{eqnarray}
where
\begin{equation}
l_L^5=\frac 1 {\sqrt{2}}\left( \begin{array}{c}
-i \nu_L \\
\nu_L\\
i \mu_L\\
\mu_L\\
0
\end{array} \right)\,,~~~~~~~~~~~~~~\mu_R^5=\left( \begin{array}{c}
0 \\
0\\
0\\
0\\
\mu_R
\end{array} \right)\,.
\end{equation}

Diagonalizing the mass matrix one finds the following expression for the muon mass
\begin{equation}
m_{\mu}\approx \frac{ f^2}{\sqrt{2}}\left[\frac {y_{L_4}y_{R_4}}{m_4}-\frac{y_{L_1}y_{R_1}}{m_1}\right] s_h c_h
\label{muonmass}
\end{equation}
valid to leading order in the mixings. 
We recall that the trigonometric dependence ($s_h\equiv \sin h/f$, $c_h\equiv \cos h/f$)
is determined by the representations of the global symmetry. One can always choose the phases so that
$m_\mu$ is real and we will assume this choice in the rest of the paper.

\subsection{Contributions to $a_\mu$}

\begin{figure}[t]
\begin{center}
\includegraphics[width=0.8\textwidth]{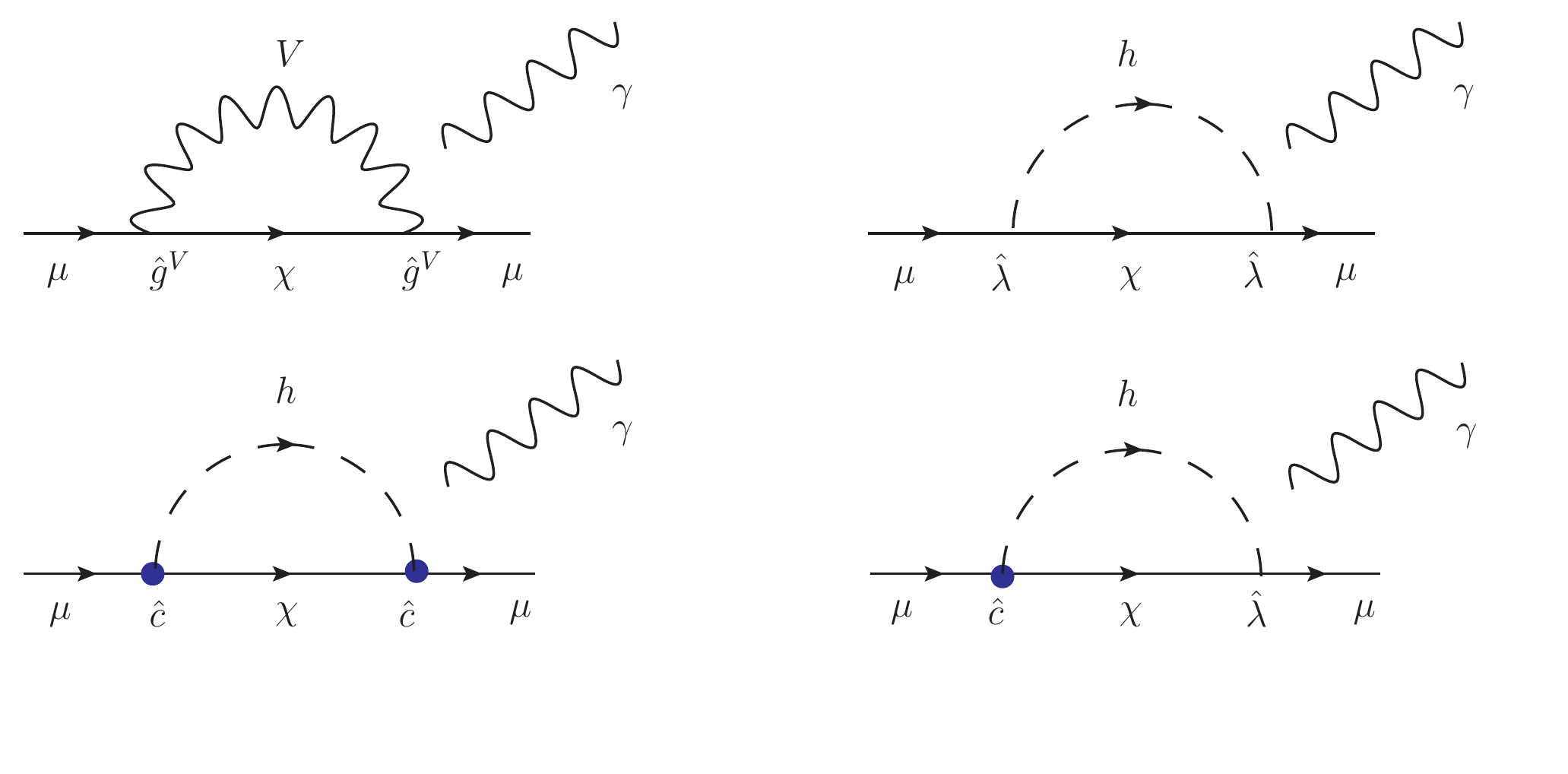}
\caption{\label{figderivative} \small Diagrams contributing to $\Delta a_\mu$. On the first line the diagrams with gauge and Yukawa interactions are shown while
on the second line the ones with Higgs derivative interactions.}
\end{center}
\label{diagrams}
\end{figure}

We parametrize the dipole moment operator of the muon as
\begin{equation}
\frac {X_\mu} {4 m_\mu} \bar{\mu}_L \sigma^{\mu\nu} \mu_R\, e F_{\mu\nu}+ h.c.
\end{equation}
For $m_\mu$ real, $a_\mu= {\rm Re}[X_\mu]$, while the imaginary part contributes to the electric dipole moment (EDM).

At 1-loop the new physics contribution to $X_\mu$  arises from diagrams with heavy fermions $\chi$  in the loop with charge -2, -1 or 0 and SM gauge fields or Higgs. To leading order $\Delta X_\mu$ is generated by diagrams with one left and one right mixing
corresponding to the function ${\cal G}$  in the expressions reported in the appendix \ref{appendixB}.

There are two classes of contributions drawn in Fig.~\ref{figderivative}. The first corresponds to diagrams with heavy composite fermions in the loop
and $W$, $Z$ or Higgs with non-derivative interactions. These are analogous to the ones considered
in renormalizable theories with vector-like fermions \cite{bib:strumia,bib:dermisek,bib:westhoff,Delaunay:2012cz,Konig:2014iqa} except that the couplings 
of the composite leptons have new contributions from the connections $e_\mu$ and $d_\mu$. 
With the standard formulas collected in the appendix A, the contribution of heavy fermions coming from this first class of diagrams reads,
\begin{eqnarray}
\label{nonderivative}
\Delta X_\mu^Z &\simeq& \frac{m_\mu m_\chi}{4\pi^2 v^2}\, (\hat{g}^Z_L)\,(\hat{g}^Z_R)^*\nonumber \\
\Delta X_\mu^{W^-}&\simeq& -\frac{m_\mu m_\chi}{8 \pi^2 v^2}\,(\hat{g}_L^{W^-})\,(\hat{g}_R^{W^-})^*\,\nonumber \\
\Delta X_\mu ^{W^+} &\simeq& \frac{m_\mu m_\chi}{8\pi^2 v^2}\, (\hat{g}_L^{W^+})\,(\hat{g}_R^{W^+})^* \nonumber\\
\Delta X_\mu^h &\simeq& \frac 1{16\pi^2}\,\frac{m_{\mu}}{m_\chi}\, (\hat{\lambda}_L)\, (\hat{\lambda}_R)^*
\end{eqnarray}
($v=246$ GeV) valid to first order in the mixings and in the limit $m_{\chi}\gg m_{Z,W,h}$. 
These contributions can be, in general, complex and generate both electric and magnetic dipole moments.
Within an explicit model the couplings in the equation above are obtained by rotating the matrices of couplings to the mass basis. 
For this purpose, given the smallness of the muon mass, it is sufficient to use the rotation matrices to first order in the mixings. 
In the CCWZ parametrization this is particularly simple since the only off-diagonal terms in the mass matrix 
are the elementary-composite mixings. The contribution from Higgs exchange is not sub-leading contrary 
to the SM where it is suppressed by $m_\mu^2/m_h^2$ compared to the gauge one. Note that in theories with vector-like leptons
without GB structure the Higgs can have additional non-derivative interactions with the heavy fermions that dominate
 \cite{Delaunay:2012cz,Konig:2014iqa}.

The second type of contribution is strictly associated to the GB nature of the Higgs and is analogous to the one considered
for dipole moment of baryons in QCD, see \cite{camalich} and Refs. therein. The term in the lagrangian (\ref{comp5}) proportional to $c_{L,R}$ contains a derivative interaction of the Higgs 
with the composite fermions. Through this vertex two new diagrams can be drawn that contribute to the dipole moment 
shown on the second line of Fig.~\ref{figderivative}. We evaluate these new contributions in the appendix B. 
The loop diagrams are finite but their values depend on the regularization procedure. 
Evaluating the integrals in 4D one finds,
\begin{eqnarray}
\Delta X_\mu^{(\partial h)^2}&\simeq& -\frac{1}{48\pi^2}\frac{m_\mu m_\chi}{f^2}\, \hat{c}_L\, \hat{c}_R^*\nonumber \\
\Delta X_\mu^{\partial h h}&\simeq& \frac{1}{24\pi^2}\frac{m_\mu}{f}\,(\hat{c}_L\,\hat{\lambda}_R^*-\hat{\lambda}_L\,\hat{c}_R^*) \  ,
\end{eqnarray}
valid within the same approximations as above.

Before analysing the explicit model above let us discuss the general structure of the result.  
The chiral structure of dipole moments is identical to the one of mass terms. As a consequence, the group theoretical structure,
controlled by the global symmetries of the theory, is also similar in the two cases. 
To leading order the dipole moment must be proportional to the product of the mixings of left and right chirality of the muon.
The Higgs dependence can be determined using a spurion analysis. To do this one should assign the elementary fields
to a representation of the global symmetry and write all the invariants under  the unbroken group using the GB matrix, see \cite{C2HDM} for more details.
One finds,
\begin{equation}
\Delta X_\mu = \sum_{A,i,j} x_A^{ij}\, y_L^i y_R^j \,(\bar{l}_L)^i U P_A^{ij} U^\dagger (\mu_R)^j 
\label{group structure}
\end{equation}
where $(l_L)^i$ and  and $(\mu_R)^j$ denote the embedding of the elementary fields into $G$ representations $\mathbf{r}_L^i$ and $\mathbf{r}_R^j$
and $P_A^{ij}$ are the projectors over the irreducible $H$ representations contained in the product of $\mathbf{r}_L^i \times \mathbf{r}_R^j$.
The coefficients $x_A^{ij}$ contain the dynamical information.

When a single invariant exists, $\Delta X_\mu$ will always be proportional to the muon mass because the Yukawa couplings have
an identical expansion as eq.~(\ref{group structure}). For the model in eqs.~(\ref{comp5}),(\ref{mixings})
this can be realised when $y_{L_4}=y_{L_1}$ and $y_{R_4}=y_{R_1}$ (other possibilities are $y_{L_1}=y_{R_1}=0$ or $y_{L_4}=y_{R_4}=0$).  
In this case one finds,
\begin{equation}
\Delta X_\mu \sim  \frac \kappa {16\pi^2}  \frac {m_\mu^2} {f^2} 
\end{equation} 
where $\kappa$ depends solely on the parameters of the composite sector 
and can be complex only if the composite sector violates CP. 
When elementary fields couple to more than one state as in (\ref{mixings})
or several invariants arise in the decomposition of $\mathbf{r}_L \times \mathbf{r}_R$, 
$\Delta X_\mu$ will not be proportional  to $m_\mu$ but will depend explicitly on the mixing parameters. 
In particular it can be complex even if the  composite sector respects CP.

\section{Results}

We now apply the tools described in the previous section to the model given by the eqs.~(\ref{comp5}),(\ref{mixings}).
The relevant couplings of the muon to the heavy fermion resonances  can be extracted from 
appendix~\ref{appendixA}. Using the formulas above we find,
\begin{eqnarray}\label{grosom2}
\Delta X_\mu&\simeq& \frac{m_{\mu}^2}{16\pi^2 f^2 }+
 \frac{m_{\mu}}{16\pi^2}\,\left[\frac 1 {\sqrt{2} m_4} y_{L_4} y_{R_4}- \frac {c_L^*} {m_1} y_{L_1} y_{R_4}- \frac {c_R} {m_1} y_{L_4} y_{R_1}+\sqrt{2}\frac{c_L^* c_R m_4}{m_1^2}  y_{L_1} y_{R_1}\right]s_h c_h \nonumber \\
&+&\frac{m_{\mu}}{24\pi^2}\,\left[\left(\frac{c_L} {m_4}-\frac{c_R} {m_1}\right)y_{L_4}y_{R_1}+  \left(\frac{c_R^*}{ m_4}-\frac{c_L^*}{ m_1}\right)y_{L_1}y_{R_4}\right]s_h c_h \nonumber\\
&+&\frac{m_{\mu}}{24\sqrt{2}\pi^2}\,\left[\frac{m_4\,c_R c_L^* }{m_1^2}y_{L_1}y_{R_1}-\frac{m_1 c_L c_R^*}{m_4^2}y_{L_4}y_{R_4}\right]s_h c_h \, 
\end{eqnarray}
to leading order in the mixings.
On the first line there are the contributions from non-derivative interactions mediated by the Higgs and the $Z$ boson respectively.
The contribution of $W$ loops is zero due to a cancellation between the diagrams with doubly charged and neutral heavy fermion in the loop.
In the second and third lines we show the contributions from the derivative Higgs interactions. 

\begin{figure}[t]
\begin{center}
\includegraphics[width=0.65\textwidth]{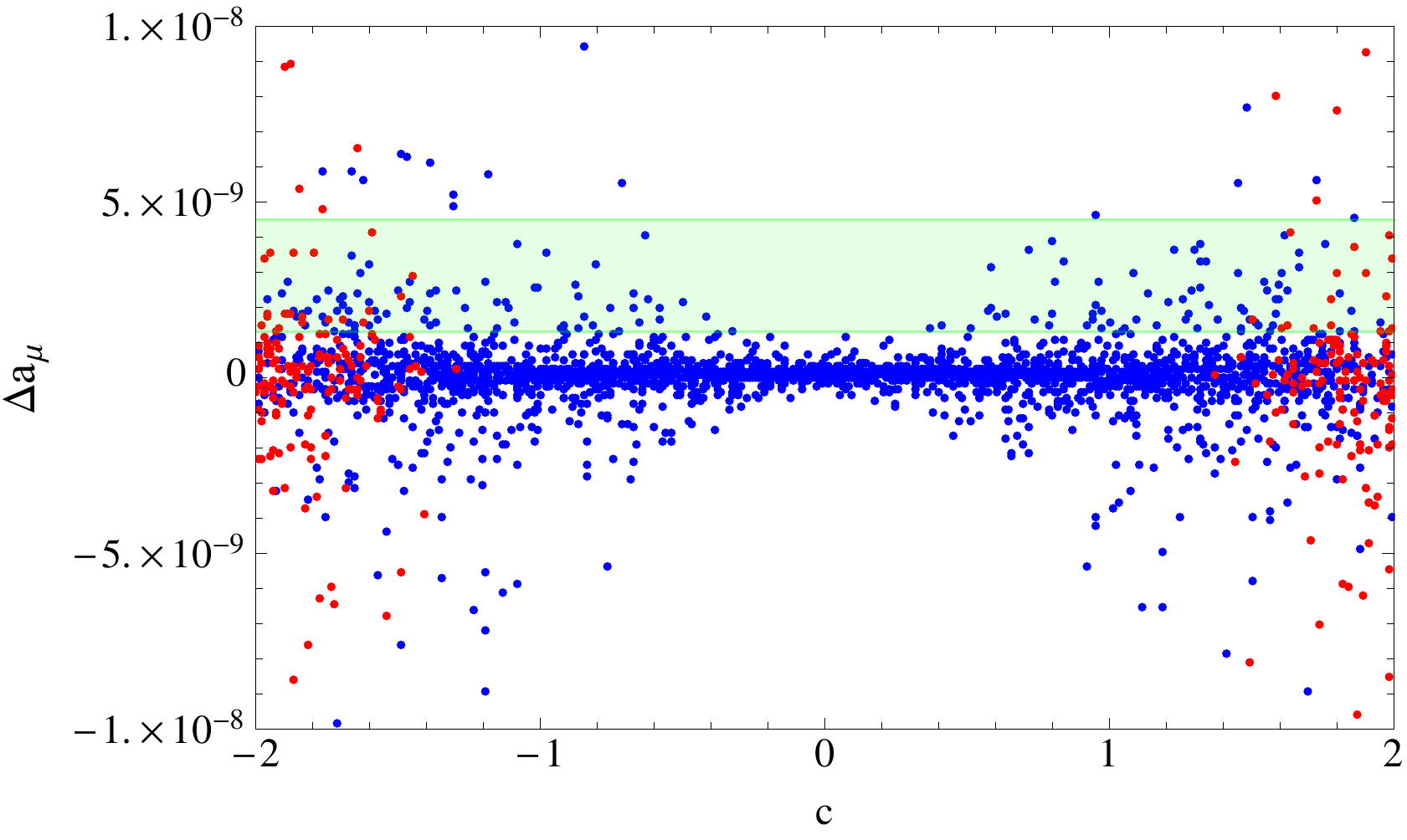}
\caption{\label{fig:4parameters} \small New physics contribution to $\Delta a_\mu$ for $c_L=c_R=c$ (real) and $f=800$ GeV. The scan is performed
by choosing $y\subset [-0.1,0.1]$ and $m_{1,4}\subset [300,3000]$ GeV. Blue points corresponds to fermionic contribution 
to the $S$ parameter $\Delta S<0.5$ assuming 3 degenerate generation partners (we use the formulas with finite terms of Ref. \cite{bib:Azatov:2013ura}). 
The green band represents the experimental value for $\Delta a_\mu$ within 2$\sigma$.}
\end{center}
\end{figure}

In Fig.~\ref{fig:4parameters} we plot a scan over the parameters of the model assuming real parameters and $c_L=c_R=c$. 
A sizable contribution to $\Delta a_\mu$  can be generated and the effect does not require extremely light fermions. 
$\Delta a_\mu$ tends to grow with $c$ but larger values of $c$ may lead to tension with bounds from $S$ parameter, see discussion below. 
We should note that $\Delta X_\mu$ is in general complex, even for real composite
sector parameters.  This implies strong bounds if a similar contribution is induced for the electron \cite{bib:passera}.

\begin{figure}[t]
\begin{center}
\includegraphics[width=0.49\textwidth]{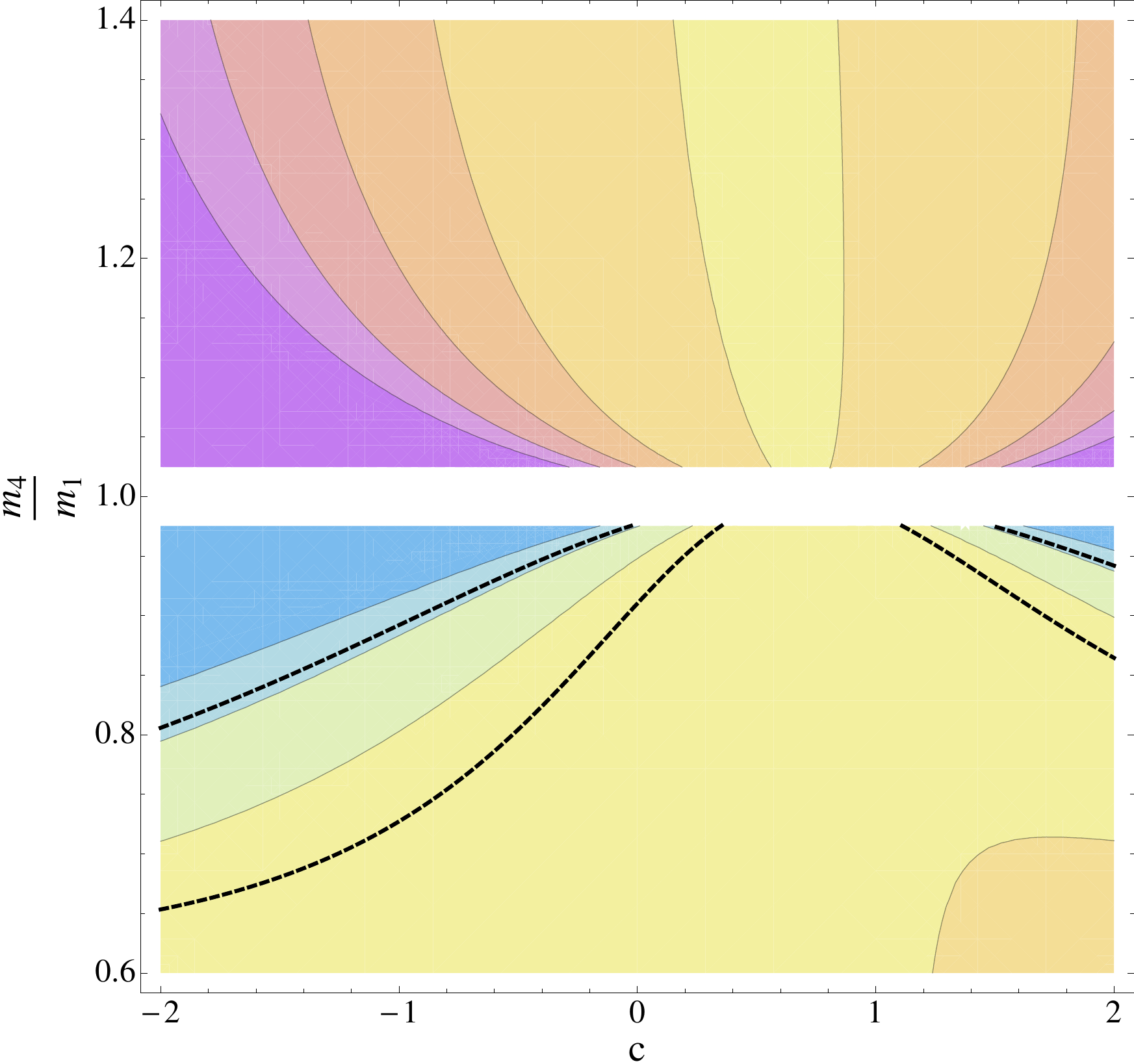}
\includegraphics[width=0.13\textwidth]{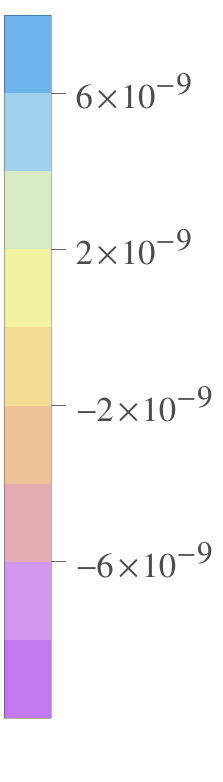}
\caption{\label{2param} \small Contribution to $\Delta a_\mu$ in the scenario with $y_{L_1}=y_{L_4}$ and $y_{R_1}=y_{R_4}$ for  $f=800$ GeV. 
The $2\sigma$ experimental value is reproduced in the region between the dashes lines. The white horizontal strip corresponds to $y_{L,R}>0.1$.}
\end{center}
\end{figure}

An interesting special case is obtained when the left and right chirality of the muon couple to a single operator of the strong sector.
This can be realized for $y_{L_1}=y_{L_4}$ and $y_{R_1}=y_{R_4}$ and it is the scenario effectively realised in 
extra-dimensional constructions (deviations form this relation correspond to non-minimal terms studied in \cite{bib:nonminimal}). 
For $c_L=c_R=c$ and real (CP and parity conserving composite sector) one finds,
\begin{equation}
\label{grosom}
\Delta X_\mu \simeq  \frac{m_{\mu}^2}{16\pi^2 f^2 }\left[1+ \frac{(m_1-\sqrt{2}\,c\, m_4)^2}{m_1 (m_1-m_4)}  
+\frac{8 }{3\sqrt{2}} c\,  -\frac{2 (m_1^2+m_1 m_4 +m_4^2)}{3 m_1 m_4}\,c^2 \right]\,.
\end{equation}
As expected $\Delta X_\mu$ is expressed in terms of the muon mass and composite sector parameters and
it is real, contributing only to the magnetic dipole moment. 
In Fig. \ref{2param} we show a contour plot of $\Delta a_\mu$ as a function of $m_4/m_1$ and $c$. 
$\Delta a_\mu$ is enhanced for a small splitting between the quadruplet and singlet masses and grows with $c$. 
The light green region gives a contribution to $\Delta a_\mu$ in agreement with the experimental value at $1\sigma$.

\subsection{Bounds}

The phenomenology of partially composite leptons was discussed in.~\cite{bib:leptonMFV}, (see also \cite{bib:westhoff}).
Due to the smallness of their masses the compositeness of SM leptons is typically small leading
to very mild constraints from modified couplings and compositeness bounds. 
For example the correction to the coupling of left-handed muons in the model discussed above is, 
\begin{equation}
\frac {\delta g_{Z\mu_L\mu_L}}{g_{Z\mu_L\mu_L}^{SM}} \simeq -\frac {v^2}{1-2 s_W^2} \left[\frac {y_{L_1}^2}{2 m_1^2}+\frac {y_{L_4}^2}{2 m_4^2}-\frac{\sqrt{2}c y_{L_1}y_{L_4}}{m_1 m_4}\right]
\end{equation}
while the coupling of $\mu_R$ does not receive corrections at tree level. Large effect can only be obtained if some chirality of leptons are strongly 
composite. The most important indirect constraint arises from the $S$ parameter. As we have seen the derivative coupling proportional to 
$c$ is a key ingredient to obtain a sizable contribution to $\Delta X_\mu$, unless the resonances are almost degenerate. The same parameter also induces a
calculable correction to $S$ from loops of composite fermions \cite{bib:panico,bib:Azatov:2013ura},
\begin{eqnarray}
\label{Sparameter}
\Delta S \simeq \frac{2}{\pi}\frac{v^2}{f^2}(1-2c^2)\log \frac{\Lambda^2}{m_4^2}+{\rm finite~terms}
\end{eqnarray}
where $\Lambda$ is an UV cutoff and finite terms depend on the regularization scheme. 
In the formula above we included a multiplicity factor for 3 generations. 
Indeed, realizing Minimal Flavor Violation (MFV) in these models requires a degenerate spectrum 
and couplings across different generations \cite{bib:compositeMFV}. In Fig.~\ref{fig:4parameters}
red points correspond to a fermionic contribution $\Delta S>$ 0.5 and are therefore disfavoured from the experimental bound. 
Other contributions to $S$ could however compensate this effect.

Direct searches from LHC exclude composite partners only up to 300-400 GeV.
The most significant difference from other models of vector-like leptons concerns Higgs couplings. 
The mass spectrum and, as a consequence, the coupling of the  Higgs to muons ($h_{\mu\mu}$) does not depend on  $c_{L,R}$,
\begin{equation}
\label{hmumu}
\frac {h_{\mu\mu}}{h_{\mu\mu}^{SM}}\simeq 1-\frac 3 2 \frac {v^2}{f^2}\,.
\end{equation}
The modification of the Higgs coupling to fermions is in fact universal to leading order, 
depending only on the representation. With a phenomenologically plausible value $f=800$ GeV or larger, $h_{\mu\mu}$ does not place a significant bound on our scenario.
This removes the correlation between $\Delta X_\mu$ and the Higgs couplings found in renormalizable models
\cite{bib:strumia,bib:dermisek}. In those Refs. the contribution to $\Delta X_\mu$ needed to reproduce the experimental anomaly 
would imply an order 5-10 modification of the decay rate of the Higgs to muons, that is on the verge of being excluded 
by LHC measurements. Moreover in a complete flavor picture realising MFV an identical modification of 
the $\tau$ coupling to the Higgs would be generated that is grossly excluded by LHC measurements.

$\Delta X_{\mu}$ in (\ref{grosom2}) is in general complex so that the imaginary part contributes 
to the muon EDM.  When only two couplings exist the phase is different from zero if the composite sector violates CP ($c_{L,R}$ complex) 
and parity ($c_L\ne c_R$). At present this does not provide a constraint for the muon but an analogous contribution 
for the electron is tightly constrained \cite{bib:passera}: the imaginary part should be suppressed by a factor $10^{-3}$
relative to $\Delta a_e$.

\section{Discussion}
\label{discussion}

In this note we computed the anomalous magnetic moment of the muon in theories with GB Higgs and partially composite fermions.
Some new features arise compared to renormalizable theories studied in the literature. In particular, interactions associated to the GB nature of the Higgs
give extra contributions that can enhance $\Delta a_\mu$ and new diagrams with Higgs derivative interactions 
exist that can give a sizable effect. Our results show that it is plausible in certain regions of parameters to obtain
a contribution that would account for the experimental anomaly. This depends crucially on the model dependent coupling 
$c$ that controls the interactions of the Higgs with the composite fermions. 

We should note that, working within a non-renormalizable effective field theory, our results should be interpreted 
as an estimate of the size of $\Delta a_\mu$ in this type of theories. Certainly we  also expect UV contributions to 
the muon magnetic moment that are uncalculable in our framework. In particular composite sector operators such as\footnote{We define
$f_{\mu \nu}\equiv U^\dagger F_{\mu\nu} U=(f_{\mu \nu}^+)^a T^a+(f_{\mu \nu}^-)^{\hat a} T^{\hat a}\equiv f_{\mu \nu}^++f_{\mu \nu}^-$.},
\begin{equation}
\frac{1}{\Lambda}\,\overline{\Psi}^i_{4L} \sigma^{\mu \nu} \Psi^j_{4R}(T^a)_{ij}(f^+_{\mu \nu})^a+h.c. \,
\label{effop}
\end{equation}
contribute to the magnetic moment of the muon. Assuming that dipoles are suppressed by a loop of the strong
dynamics (as for example in weakly coupled 5D realizations of our framework) we find that their typical size is,
\begin{equation}
\Delta a_\mu^{UV}\sim \frac 1 {16\pi^2}  \frac {m_\mu^2}{f^2}\,
\end{equation}
which is an order of magnitude smaller than required to reproduce the anomaly for $f=800$ GeV. The IR contribution from loops of light degrees
of freedom would be in this case dominant. Nevertheless, we cannot a priori exclude that larger UV contributions are present.

It is interesting to cast our results into the broader flavor picture of partially composite Higgs models, see \cite{bib:leptonMFV,bib:compositeMFV}
for a detailed discussion. The hypothesis of partial compositeness can suppress flavor transitions beyond the SM. 
Nevertheless, severe bounds exist especially in the lepton sector. For example ${\rm Br}[\mu\to e \gamma]< 5 \times10^{-13}$
hints to a scale of compositeness $\Lambda > 50 $ TeV much larger than the value expected for these models if they are relevant to
the hierarchy problem. Tension with flavor constraints can be eliminated if the theory realizes MFV. In fact, partial compositeness
allows to elegantly realize this hypothesis: this requires that the composite sector possesses flavor symmetries 
that are only broken by mixings proportional to the SM Yukawa couplings.  This can be realized if left-handed or right-handed fermions 
have equal degree of compositeness. One interesting prediction is that the contribution to the 
$(g-2)$ of the electron is related to the one of the muon as,
\begin{equation}
\frac {\Delta a_e}{\Delta a_\mu}=\frac {m_e^2}{m_\mu^2}
\end{equation}
that could be of interest in future experiments \cite{bib:passera}.
Moreover contributions to EDMs are automatically zero at 1-loop if the strong sector also respects CP.

Our results can be extended in various directions. First, models with different representations of composite fermions or
different patterns of symmetry breaking can be studied with the techniques described in this paper and other dipole moments 
relevant for composite Higgs models can be computed. One obvious generalization is for example  the computation of chromo-magnetic operators in the quark sector. 
The same type of effects studied here also appears in models with extra-dimensions that correspond to an infinite number of resonances with derivative couplings 
determined by the metric.  Finally the contribution of composite spin-1 resonances could also be studied along the lines described in this paper.

\vspace{1cm}

\subsection*{Acknowledgments}
The work of OA and MR is supported by the MIUR-FIRB grant RBFR12H1MW. 
MR would like to thank Roberto Contino and Giuliano Panico for discussions.

\vspace{1.5cm}

\appendix

\section{Relevant Formulas}
\label{appendixA}

In the CCWZ formalism one introduces the GB matrix,
\begin{equation}
U=e^{i \frac{ \sqrt{2}}f \pi^{\hat a} T^{\hat a}}
\end{equation}
where $T^{\hat a}$ are the broken generators, and constructs the Maurer-Cartan form
\begin{equation}
U^\dagger[A_\mu+i\partial_\mu]U=i\, U^\dagger D_\mu U= i\,d_\mu^{\hat{a}}T^{\hat{a}}+i\,e_\mu^aT^a\,.
\end{equation} 
Explicitly for $SO(5)/SO(4)$ this is given by,
\begin{eqnarray}\label{d}
d^{\hat{a}}_\mu &=& \sqrt{2}\left(\frac{1}{f}-\frac{\sin \pi/f}{\pi}\right)\frac{\vec{\pi}\cdot D_\mu \vec{\pi}}{\pi^2}{\, \pi^{\hat{a}}}+\sqrt{2}\,\frac{\sin \pi/f}{\pi}D_\mu \pi^{\hat{a}}\nonumber \\
e_\mu^a &=&-A_\mu^a +4\,i\,\frac{\sin^2(\pi /2f)}{\pi^2}\,\vec{\pi}^Tt^aD_\mu\vec{\pi}\label{e}
\end{eqnarray}
with $t^a$ the $SO(4)$ generators in  4x4 matrix form and
\begin{equation}
D_\mu \pi^{\hat{a}}=\partial_\mu \pi^{\hat{a}} -i\,A^a_\mu\,(t^a)^{\hat{a}}_{\,\,\,\hat{b}}\, \pi^{\hat{b}}\,\,.
\end{equation}

For the model in section \ref{sec:2} the lagrangian can be written explicitly as,
\begin{eqnarray}
\label{mass}
\mathcal{L}&=&{\cal L}_{kinetic}-(\,\overline{\Theta}_L\mathcal{M}_{-1}\Theta_R + \overline{\mathcal{N}}_L\mathcal{M}_{N}\,N_R+ h.c.)-\,m_4\overline{E}_{-2}E_{-2}\nonumber \\
&+&\frac{g}{\sqrt{2}}\left[\overline{\mathcal{N}}\,_L\,g_L^{WN}\,\displaystyle{\not} W^+\Theta_L + \overline{N}_R\,g_R^{WN}\,\displaystyle{\not} W^+\Theta_R + \overline{E}_{-2L}\,g_L^{WC}\,\displaystyle{\not} W^-\Theta_L +\overline{E}_{-2R}\,g_R^{WC}\,\displaystyle{\not} W^-\Theta_R +h.c. \right]  \nonumber \\
&+&\frac{g}{c_W}\,\left[\overline{\Theta}\,_L\,g_L^{Z}\,\displaystyle{\not} Z\,\Theta_L + \overline{\Theta}\,^T_R\,g_R^Z\,\displaystyle{\not} Z\,\Theta_R\right] +
 i\,\frac{c_L}f\,\overline{\Theta}_L {\cal R} \, \displaystyle{\not} \partial h\,\Theta_L+ i\,\frac{c_R}f \,\overline{\Theta}_R {\cal R}\, \displaystyle{\not} \partial h\,\Theta_R
\end{eqnarray}
where we have defined the fields,
\begin{equation}
\Theta_{L,R} = \left( \begin{array}{c}
\mu \\
E\\
E_{-1}\\
\tilde{E}
\end{array} \right)_{L,R} \qquad
\mathcal{N}_{L} = \left( \begin{array}{c}
\nu \\
N\end{array} \right)_L
\end{equation}
the mass matrices
\begin{equation}
\mathcal{M}_{-1}  =
\left( \begin{array}{cccc}
0 & \displaystyle y_{L_4} f\, \frac{1+c_h}{2}\, \,\,& \displaystyle y_{L_4}f\, \frac{1-c_h}{2}\,\,\, & \displaystyle y_{L_1} f\, \frac{s_h}{\sqrt{2}} \\
-\displaystyle y_{R_4}f\, \frac{s_h}{\sqrt{2}} & m_4 & 0 & 0 \\
\displaystyle y_{R_4}f\, \frac{s_h}{\sqrt{2}} & 0 & m_4 & 0\\
y_{R_1}\,f\,c_h & 0 & 0 & m_1 \end{array} \right)
\qquad
\mathcal{M}_{N}  =
\left( \begin{array}{c}
y_{L_4} f \\
m_4
\end{array} \right)\,
\label{massmatrix1}
\end{equation}
and the couplings 
\begin{eqnarray}
g^Z_L&=& 
\left( \begin{array}{cccc}
-\frac{1}{2}+s^2_{W} & 0 &0&0 \\
0 & - \displaystyle \frac{c_h}{2}+s^2_W & 0& -\displaystyle c_L\,\frac{s_h}{2} \\
0 & 0 &\displaystyle \frac{c_h}{2}+s^2_{W}& -\displaystyle c_L\,\frac{s_h}{2}\\
0&-\displaystyle c_L^*\,\frac{s_h}{2}&-\displaystyle c_L^*\,\frac{s_h}{2}&s^2_{W}
\end{array} \right) \nonumber \\
g^Z_R&=& 
\left( \begin{array}{cccc}
+s^2_W & 0 &0&0 \\
0 & - \displaystyle \frac{c_h}{2}+s^2_{W} & 0& -\displaystyle c_R\,\frac{s_h}{2} \\
0 & 0 &\displaystyle \frac{c_h}{2}+s^2_{W}& -\displaystyle c_R\,\frac{s_h}{2}\\
0&-\displaystyle c_R^*\,\frac{s_h}{2}&-\displaystyle c_R^*\,\frac{s_h}{2}&s^2_{W}
\end{array} \right) \nonumber \\
 g_L^{WN}&=&\left( \begin{array}{cccc}
1 & 0 & 0&0 \\
0 & \displaystyle \frac{1+c_h}{2} & \displaystyle \frac{1-c_h}{2}  & c_L\,s_h
\end{array} \right)\qquad  g_R^{WN}=\left( \begin{array}{cccc}
0 & \displaystyle \frac{1+c_h}{2} & \displaystyle \frac{1-c_h}{2}  & c_R\,s_h \\
\end{array} \right)\nonumber \\
g_L^{WC}&=&\left( \begin{array}{cccc}
0 & \displaystyle \frac{1-c_h}{2} & \displaystyle  \frac{1+c_h}{2}  & -c_L\,s_h
\end{array} \right)\,\,\,\,\,\,\,\,  g_R^{WC}=\left( \begin{array}{cccc}
0 & \displaystyle  \frac{1-c_h}{2} & \displaystyle  \frac{1+c_h}{2}  & -c_R\,s_h 
\end{array} \right) \nonumber \\
{\cal R} &=&\left(\begin{array}{cccc}
0 & 0 & 0 & 0  \\
0 & 0& 0 & 1 \\
0 & 0 & 0 & -1 \\
0 &-1 & 1 & 0 
\end{array}\right)\,.
\end{eqnarray} 
The relevant couplings used in the paper (denoted with a "hat") are obtained rotating to the physical mass basis defined by Eq.~(\ref{massmatrix1})
(Higgs Yukawa couplings are given by $\lambda =d\,\mathcal{M}_{-1}/d\langle h \rangle$).
Explicit formulae are easily derived to first order in the mixings sufficient for the analysis in this paper.



\section{Dipole Moments}
\label{appendixB}

In this appendix we present the relevant formulas for  dipole moments in theories with GB Higgs.
At 1-loop only states with charge -2, -1, 0 ($\chi_{-2,-1,0}$) contribute. We consider the following interaction terms,
\begin{eqnarray}
{\cal L}_{int} &=&[ V_0^\mu g^{V_0}_L \bar\mu_L \gamma_\mu \chi_{-1L} +V_+^\mu g^{V_+}_L \bar\mu_L \gamma_\mu \chi_{-2L}+  V_-^\mu g^{V_-}_L \bar\mu_L \gamma_\mu \chi_{0L} \nonumber \\
&-&\lambda_L \bar{\mu}_L\, h\, \chi_R +i \frac{C_L}f  \bar{\mu}_L\,  \displaystyle{\not} \partial h \, \chi_L +(L\rightarrow R)]+h.c.
\end{eqnarray}

\subsection{Non-derivative Interactions}

With the couplings on the first line one finds the following contributions to the muon magnetic moment,
\begin{eqnarray}\label{eq:v0}
\Delta X_\mu^{V_0}&=&\frac{m_\mu^2}{8\,\pi^2m_{V_0}^2}{\left[(|g_L^{V_0}|^2+|g_R^{V_0}|^2)\,\mathcal{F}_{V_0}(x)+g_L^{V_0} (g_R^{V_0})^*\,\mathcal{G}_{V_0}(x)\frac{m_\chi}{m_\mu}\right]}\nonumber\\
\Delta X _\mu^{V_-} &=&\frac{m_\mu^2}{16\,\pi^2m_{V_-}^2}{\left[(|g_L^{V_-}|^2+|g_R^{V_-}|^2)\,\mathcal{F}_{V_-}(x)+g_L^{V_-} (g_R^{V_-})^*\,\mathcal{G}_{V_-}(x)\frac{m_\chi}{m_\mu}\right ]}\nonumber\\
\Delta X _\mu^{V_+} &=& \frac{m_\mu^2}{16\,\pi^2m_{V_+}^2}{\left[(|g_L^{V_+}|^2+|g_R^{V_+}|^2)\,(4 \mathcal{F}_{V_0}(x)+\mathcal{F}_{V_-}(x))+g_L^{V_+} (g_R^{V_+})^*\,(4\mathcal{G}_{V_0}(x)+\mathcal{G}_{V_-}(x))\frac{m_\chi}{m_\mu}\right ]} \nonumber\\
\Delta X_\mu^h&=& \frac{m_\mu^2}{16\,\pi^2 m_h^2}{\left[(|\lambda_L|^2+|\lambda_R|^2)\,\mathcal{F}_h(x)+\lambda_L \lambda_R^*\,\mathcal{G}_h(x)\frac{m_\chi}{m_\mu}\right ]}
\end{eqnarray}
respectively for diagrams with $V^0$, $V^\pm$ and $h$ in the loop. Here $m_\chi$ the mass of the heavy fermion. The loop functions are given by
\begin{eqnarray}
\label{fv0}
\mathcal{F}_{V_0}(x)&=&\frac{-5x^4+14x^3+18x^2\,\log x-39x^2+38x-8}{12(x-1)^4}\\
\mathcal{G}_{V_0}(x)&=&\frac{x^3-6x\,\log x+3x-4}{2(x-1)^3}\\
\mathcal{F}_{V_-}(x)&=&\frac{4x^4+18x^3\,\log x-49x^3+78x^2-43x+10}{6(x-1)^4}\\
\mathcal{G}_{V_-}(x)&=&\frac{-x^3-6x^2\,\log x+12x^2-15x+4}{(x-1)^3}\\
\mathcal{F}_h(x)&=&\frac{x^3-6 x^2+6 x\,\log x+3 x+2}{6 (x-1)^4}\\
\mathcal{G}_h(x)&=&\frac{x^2-4 x+2 \log x+3}{ (x-1)^3}
\label{gv}
\end{eqnarray}
with $x=m_{\chi}^2/ m_{V_0,V_+,h}^2 $.\\

\subsection{Derivative Interactions}

The contribution of the diagram with two Higgs derivative interactions is formally given by,

\begin{eqnarray}
\Delta X_\mu^{(\partial h)^2}&\sim&\int_0^1 u du\int \frac {d^4l}{(2\pi)^4} \frac {A\, l^2+ B}{(l^2-\Delta)^3}\nonumber \\
A&=& (2-3 u)\left[m_\mu^2 (|C_L|^2+|C_R|^2)-m_\mu m_\chi C_L C_R^*\right]\nonumber \\
B&=& 2\,\{\,[\,m_\mu^4\,(u^2-u^3)-m_\mu^2\,m_\chi^2\,u^2\,]\,(|C_L|^2+|C_R|^2)-\,m_\mu^3\,m_\chi\,u^3\,C_L C_R^*\,\} \nonumber \\
\Delta&=& u(u-1)m_{\mu}^2+(1-u)m_h^2+u m_{\chi}^2
\end{eqnarray}

Naively the integral over momenta is logarithmically divergent and needs to be regularized.
One can see that upon integration over $u$ the result is finite but it depends on the regulator chosen. 
The different results correspond to the addition of UV local operators such us (\ref{effop}) to the effective action. 
For our estimates we perform the integral in 4D. Neglecting the muon mass relative to $m_h$ and $m_\chi$ we find,
\begin{eqnarray}
\Delta X_\mu^{(\partial h)^2}&=&{ - \frac{m_\mu^2}{16\,\pi^2 f^2}\left[(|C_L|^2+|C_R|^2)\,\mathcal{F}_{(\partial h)^2}(x)+C_L C_R^* \,\mathcal{G}_{(\partial h)^2}(x)\frac{m_\chi}{m_\mu}\right]}\nonumber \\
\mathcal{F}_{(\partial h)^2}(x)&=&{\frac{-2 x^4-12 x^3+6 \left(2 x-1\right) x^2 \log x+27 x^2-16 x+3}{6 \left(x-1\right)^4}}\nonumber\\
&{ +}& {\frac{m_\mu^2}{m_\chi^2}\,\frac{3 x^4+\left(24 x^4-12 x^3\right) \log x+10 x^3-18 x^2+6 x-1}{12 \left(x-1\right)^5}} \nonumber \\
\mathcal{G}_{(\partial h)^2}(x)&=&{\frac{2 x^3-6 x^2 \log x+3 x^2-6 x+1}{3 \left(x-1\right)^3}}\nonumber\\
&{-}&{\frac{m_\mu^2}{m_\chi^2}\,\frac{2 x^3-6 x^2 \log x+3 x^2-6 x+1}{6 \left(x-1\right)^3}}
\end{eqnarray}

The diagram with one derivative interaction and a Yukawa coupling has very similar features. 
In this case one finds,
\begin{eqnarray}
\Delta X_\mu^{\partial h h}&=& {- \frac{m_\mu}{16\,\pi^2 f}[(C_L^*\lambda_L+C_R\lambda_R^*)\,\mathcal{F}_{\partial h h}(x)+
(C_L \lambda_R^*-\lambda_L C_R^*)\,\mathcal{G}_{\partial h h}(x)]}\nonumber \\
\mathcal{F}_{\partial h h}(x)&=&{\frac{m_{\mu}}{m_{\chi}}\,\frac{6 x^3 \log x-11 x^3+18 x^2-9 x+2}{3 \left(x-1\right)^4}}\nonumber \\
\mathcal{G}_{\partial h h}(x)&=& {\frac{x^3-6 x^2 \log x+6 x^2-9 x+2}{3 \left(x-1\right)^3}}\nonumber \\
&+& {\frac{m_\mu^2}{m_\chi^2}\,\frac{7 x^4+12 \left(2 x^4-2 x^3+x^2\right) \log x+12 x^3-36 x^2+20 x-3}{12 \left(x-1\right)^5}}
\end{eqnarray}

\end{document}